\documentclass[twocolumn,prl,showpacs,aps]{revtex4-1}

\usepackage[english]{babel}
\usepackage[ansinew]{inputenc}
\usepackage{times}
\usepackage{graphicx}
\usepackage{graphics}
\usepackage{amsmath}
\usepackage{amsfonts}
\usepackage{amssymb}
\usepackage{dsfont}
\usepackage{epstopdf}
\usepackage{makeidx}
\usepackage{subfigure}
\usepackage{color}
\usepackage{pgf}
\usepackage{bm}
\usepackage{tikz} 
\usepackage[normalem]{ulem}

\newcommand{\be}{\begin{equation}}
\newcommand{\ee}{\end{equation}}
\newcommand{\bea}{\begin{eqnarray}}
\newcommand{\eea}{\end{eqnarray}}

\makeindex
\begin{document}
\title{Tuning plasmonic cloaks with an external magnetic field}
\author{W. J. M.  Kort-Kamp}
\affiliation{Instituto de F\'{\i}sica, Universidade Federal do Rio de Janeiro,
Caixa Postal 68528, Rio de Janeiro 21941-972, RJ, Brazil}
\author{F. S. S. Rosa}
\affiliation{Instituto de F\'{\i}sica, Universidade Federal do Rio de Janeiro,
Caixa Postal 68528, Rio de Janeiro 21941-972, RJ, Brazil}
\author{F. A. Pinheiro}
\affiliation{Instituto de F\'{\i}sica, Universidade Federal do Rio de Janeiro,
Caixa Postal 68528, Rio de Janeiro 21941-972, RJ, Brazil}
\author{C. Farina}
\affiliation{Instituto de F\'{\i}sica, Universidade Federal do Rio de Janeiro,
Caixa Postal 68528, Rio de Janeiro 21941-972, RJ, Brazil}

\date{\today}

\begin{abstract}

We propose a mechanism to actively tune the operation of plasmonic cloaks with an external magnetic field by investigating electromagnetic scattering by a dielectric cylinder coated with a magneto-optical shell. In the long wavelength limit we show that the presence of a magnetic field may drastically reduce the scattering cross-section at all observation angles.
We demonstrate that the application of magnetic fields can modify the operation wavelength without the need of changing material and/or geometrical parameters. We also show that applied magnetic fields can reversibly switch on and off the cloak operation. These results, which could be achieved for existing magneto-optical materials, are shown to be robust to material losses, so that they may pave the way for developing actively tunable, versatile plasmonic cloaks.

\end{abstract}
\maketitle
%

The idea of rendering an object invisible in free space, which had been restricted to human imagination for many years, has become an important scientific and technological endeavour since the advent of metamaterials. Progress in micro- and nanofabrication have fostered the construction of artificial metamaterials with unusual electromagnetic (EM) properties~\cite{zheludev2012}, which have been exploited in the development of a variety of approaches for designing EM cloaks and achieving invisibility. Among these approaches one can highlight the coordinate-transformation method~\cite{pendry2006,leonhardt2006,schurig2006} and the scattering cancelation techniques~\cite{alu2005,alu2008,edwards2009,filonov2012,chen2012,rainwater2012,KortKamp2013,nicorovici1994}. The coordinate-transformation method, grounded in the emerging field of transformation optics~\cite{chen2010}, requires metamaterials with anisotropic and inhomogeneous profiles, which are able to bend the incoming EM radiation around a given region of space, rendering it invisible to an external observer. This method has been first experimentally realized for microwaves~\cite{schurig2006}, and later extended to infrared and visible frequencies~\cite{valentine2009,ergin2010}. The scattering cancellation technique, which constitutes the basis for the development of plasmonic cloaks, was proposed in~\cite{alu2005}. Applying this technique, a dielectric or conducting object can be effectively cloaked by covering it with a homogeneous and isotropic layer of plasmonic material with low-positive or negative electric permittivity. In these systems, the incident radiation induces a local polarization vector in the shell that is out-of-phase with respect to the local electric field so that the in-phase contribution given by the cloaked object may be partially or totally canceled~\cite{alu2005,alu2008,chen2012}. Experimental realizations of cylindrical plasmonic cloaks for microwaves exist in 2D~\cite{edwards2009} and 3D~\cite{rainwater2012}, paving the way for many applications in camouflaging, low-noise measurements and non-invasive sensing~\cite{alu2008,chen2012}.

Despite the notable progresses in all cloaking techniques, the development of a practical, versatile cloaking device remains a challenge. One of the reasons is that the operation frequency of many cloaking devices is typically narrow, although some photonic crystals with large, complete photonic band gaps exist and could be explored in the design of wide band cloaking systems~\cite{lin1}. Another reason is that the majority of cloaking devices developed so far are generally tailored to operate at a given frequency band, that cannot be freely modified after fabrication. As a result, if there is a need to modify the operation frequency band it is usually necessary to engineer a new device, with different geometric parameters and materials, limiting its applicability~\cite{alu2008}. To circumvent this limitation, some proposals of tunable cloaks have been developed \cite{peining,zharova2012,milton2009}. One of them is to introduce a nonlinear layer in multi-shell plasmonic cloaks, so that the scattering cross-section can be controlled by changing the intensity of the incident EM field~\cite{zharova2012}. Core-shell nonlinear plasmonic particles can also be designed to exhibit Fano resonances, allowing a swift switch from being completely cloaked to being strongly resonant~\cite{monticone2013}. Another implementation of tunable plasmonic cloaks involves the use of a graphene shell~\cite{chen2011}.
However, most of these proposals are based either on a {\em passive} mechanism of tuning the cloaking device ({\em e.g.}, by chemically modifying the graphene surface by carboxylation and thiolation~\cite{chen2011,chuang2007}) or depends on a given range of intensities for the incident excitation, as in the case of nonlinear plasmonic cloaks~\cite{zharova2012}.

Here we propose an alternative mechanism to {\em actively} tune the operation of plasmonic cloaks with an external magnetic field ${\bf B}$. We show that this is feasible by investigating EM scattering by a dielectric cylinder coated by a magneto-optical shell. The application of ${\bf B}$ drastically reduces the differential scattering cross-section; this reduction can be as high as $95\%$ if compared to the case without ${\bf B}$. The presence of ${\bf B}$ modifies the operation wavelength without the need of changing material and/or geometrical parameters. Besides, ${\bf B}$ could be used as an external agent to reversibly switch on and off the operation of the cloak. Our results are robust to material losses and could be achieved for existing magneto-optical materials and moderate magnetic fields, so that they might be useful for developing actively tunable, versatile plasmonic cloaks.

\begin{figure}
  \centering
  \includegraphics[scale=0.3]{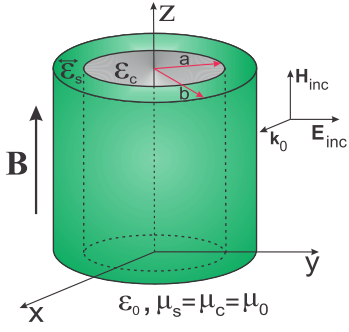}
  \caption{Schematic representation of the scattering system.}\label{Figura1}
\end{figure}

In Fig.~\ref{Figura1} we show the scheme of the system: an infinitely long, non-magnetic cylinder with permittivity $\varepsilon_c$ and radius $a$ coated with a magneto-optically active cylindrical shell with magnetic permeability $\mu_0$ and radius $b>a$. The system is subjected to a uniform ${\bf B}$ parallel to the cylindrical symmetry axis. Let us consider a TM plane wave of frequency $\omega$ impinging normally on the cylinder. The background medium is vacuum. In the presence of ${\bf B}$ the shell permittivity $\stackrel{\leftrightarrow}{\varepsilon}_s$ is anisotropic; for the geometry of Fig.~\ref{Figura1} it reads~\cite{Stroud1990}
\begin{eqnarray}
\label{TensoresEeM}
\stackrel{\leftrightarrow}{\varepsilon}_s =
\left(
  \begin{array}{ccc}
    \varepsilon_{xx} & \varepsilon_{xy} & 0 \\
    \varepsilon_{yx} & \varepsilon_{yy} & 0 \\
    0 & 0 & \varepsilon_{zz} \\
  \end{array}
\right)
= \left(
  \begin{array}{ccc}
    \varepsilon_{s} & i\gamma_s & 0 \\
    -i\gamma_s & \varepsilon_{s} & 0 \\
    0 & 0 & \varepsilon_{zz}, \\
  \end{array}
\right)
\end{eqnarray}
where the off-diagonal term $\gamma_s$ has a dispersive character and depends upon $B$, vanishing in its absence. EM scattering from a coated cylinder can be analyzed with Mie theory~\cite{BohrenHuffman, Stroud1990}. The scattering cross-section efficiency $Q_{\textrm{sc}}$  is
\begin{equation}
\label{Qsc}
Q_{\textrm{sc}} = \dfrac{2}{k_0b}\,\displaystyle{\sum_{-\infty}^{+\infty}|D_m|^2},
\end{equation}
where $D_m$ are the scattering coefficients and $k_0 = 2\pi/\lambda$ is the incident wavenumber. We consider the dipole approximation, i.e. $k_0b\ll 1\, , \ k_cb\ll 1\, , \ k_sb\ll1$, where the dominant scattering terms are $m = 0\, , \ m = \pm 1$, and $k_c$ ($k_s$) is the core (shell) wavenumber. The condition for invisibility is determined by imposing that each scattering coefficient vanishes separately.  In the dipole approximation, $D_0$ is identically zero for non-magnetic core and shell. For $D_{-1}$ and $D_{+1}$ to vanish, the ratio $\eta \equiv a/b$ must be
\begin{equation}
\label{razaoraios}
\eta_{\pm 1} = \sqrt{\dfrac{(\varepsilon_0\pm \gamma_s-\varepsilon_s)(\varepsilon_c \pm \gamma_s+\varepsilon_s)}
{(\varepsilon_c \pm \gamma_s-\varepsilon_s)(\varepsilon_0 \pm \gamma_s+\varepsilon_s)}},
\end{equation}
where $\eta_{-1}$ and $\eta_{+1}$ represent the values of $\eta$ that vanish $D_{-1}$ and $D_{+1}$, respectively.

\begin{figure}
  \centering
  \includegraphics[scale=0.42]{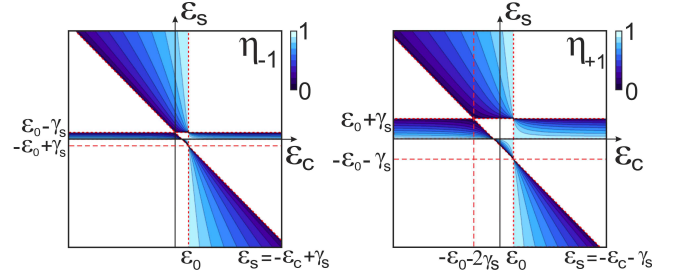}
  \caption{Regions for which the invisibility condition (\ref{razaoraios}) is satisfied for the corresponding $\eta_{\pm1}$ between 0 and 1 in the presence of off-diagonal terms in electric permittivity, proportional to the applied magnetic field.}\label{Figura2}
\end{figure}

The conditions for achieving the cancellation of $D_{-1}$ and $D_{+1}$, obtained from Eq.~(\ref{razaoraios}), are shown in Fig.~\ref{Figura2}, as a function of the permittivities of the inner core and of the outer cloaking shell. In Fig.~\ref{Figura2}, blue regions correspond to situations where either $D_{-1}$ or $D_{+1}$ vanish for a given $\gamma_s$ and different ratios $\eta_{\pm1}$ ($0\leq \eta_{\pm1} \leq 1$). From Fig.~\ref{Figura2} and Eq.~(\ref{razaoraios}) we conclude that it is not possible to obtain $\eta_{-1} = \eta_{+1}$ for $\gamma_s \neq 0$, {\em i.e.} the contributions of $D_{-1}$ and $D_{+1}$ to $Q_{sc}$ cannot be simultaneously canceled for the same $\eta$ in the presence of ${\bf B}$. However, it is still possible to drastically reduce $Q_{sc}$ and, for a given $\eta$, obtain values for the scattering cross-section that are substantially smaller in the presence of ${\bf B}$. Indeed, guided by Fig.~\ref{Figura2} and by inspecting Eq.~(\ref{razaoraios}), one can choose a ratio $\eta_{-1} $ or $\eta_{+1}$ that vanishes the contribution of either $D_{-1}$ or $D_{+1}$ to $Q_{sc}$, respectively. As a result, a strong reduction of $Q_{sc}$ can be achieved since $D_{0}$ is already zero for non-magnetic materials.

\begin{figure}
\centering
\includegraphics[scale=0.42]{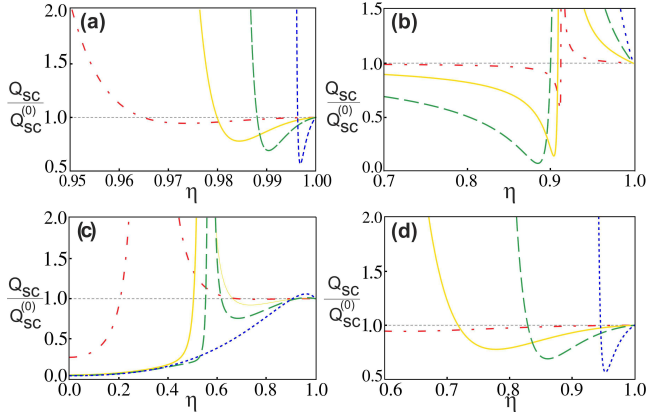}
\caption{Scattering efficiency $Q_{sc}$ (normalized by its value in the absence of ${\bf B}$, $Q_{sc}^{(0)}$) as a function of the ratio between the external and internal radii $\eta$ for (a) $\varepsilon_c = 10\varepsilon_0$ and $\varepsilon_s = 0.1 \varepsilon_0$, (b) $\varepsilon_c = 0.1\varepsilon_0$ and $\varepsilon_s = 10\varepsilon_0$, (c) $\varepsilon_c = 10\varepsilon_0$ and $\varepsilon_s = -1.5\varepsilon_0$, and (d) $\varepsilon_c = 10\varepsilon_0$ and $\varepsilon_s = 1.5\varepsilon_0$. Different curves correspond to different magnitudes of the off-diagonal term of the electric permittivity $\gamma_s$, proportional to $B$: $\gamma_s = \varepsilon_0$ (dash-dotted curve), $\gamma_s = 3\varepsilon_0$ (solid curve), $\gamma_s = 5\varepsilon_0$ (dashed curve), $\gamma_s = 12\varepsilon_0$ (dotted curve).}\label{Figura3}
\end{figure}

In Fig.~\ref{Figura3}, $Q_{sc}$ in the presence of ${\bf B}$ ($\gamma_s \neq 0$) is calculated as a function of $\eta$ for different $\varepsilon_c$ and $\varepsilon_s$.
$Q_{sc}$ is normalized by the scattering efficiency in the absence of ${\bf B}$, $Q_{sc}^{(0)}$, and $b=0.01 \lambda$ (i.e. dipole approximation is valid). Figure~\ref{Figura3}a corresponds to the situation where, given the parameters $\varepsilon_c = 10\varepsilon_0$ and $\varepsilon_s = 0.1 \varepsilon_0$, it is possible to achieve invisibility for $\eta \simeq 0.92$ in the absence of the magnetic field \cite{alu2005}. To see this it suffices to put $\gamma_s =  0$ and substitute the set of material parameters used in Fig.~\ref{Figura3}a in Eq.~(\ref{razaoraios}). The application of ${\bf B}$ shifts the values of $\eta$ that cause a strong reduction of $Q_{sc}$ to higher values. 
Interestingly, by increasing ${\bf B}$ one can not only shift the operation range of the device for $\eta$ close to 1, but also significantly decrease $Q_{sc}$. Indeed, for $0.98 \leq \eta < 1$ the application of ${\bf B}$ leads to a reduction of the order of 50\% in $Q_{sc}$ if compared to the case where ${\bf B} = {\bf 0}$. This result indicates that, for this set of parameters, an optimal performance of the cloak could be achieved for very thin magneto-optical films.

In Fig.~\ref{Figura3}b, for the chosen set of parameters ($\varepsilon_c = 0.1\varepsilon_0$ and $\varepsilon_s = 10\varepsilon_0$) EM transparency also occurs for $\eta \simeq 0.92$ in the absence of ${\bf B}$. In the presence of ${\bf B}$ achieving perfect invisibility (vanishing scattering cross-section) is no longer possible for $\eta \simeq 0.92$, since a peak in $Q_{sc}$ emerges precisely for this value of $\eta$. Physically, this peak in $Q_{sc}$ can be explained by the fact that the EM response of the shell is no longer isotropic in the presence of ${\bf B}$; hence the induced electric polarization within the shell is not totally out-of-phase with respect to the one induced within the inner cylinder. The net effect is that these two induced electric polarizations are not capable to cancel each other anymore in the presence of ${\bf B}$, resulting in a non-vanishing scattered field. The fact that the application of ${\bf B}$ induces a peak in $Q_{sc}$ in a system originally conceived to achieve perfect invisibility (in the absence of ${\bf B}$) suggests that a magnetic field could be used as an external agent to reversibly switch on and off the operation of the cloak. On the other hand, for $\eta \lesssim 0.9$ and this set of parameters the application of ${\bf B}$ always lead to a reduction of $Q_{sc}$ with respect to $Q_{sc}^{(0)}$, increasing the efficiency of the cloak. For example, for $\gamma_s = 5\varepsilon_0$ and $\eta = 0.88$ it is possible to achieve a reduction of $Q_{sc}$ of the order of 93\% with respect to $Q_{sc}^{(0)}$; for $\gamma_s = 3\varepsilon_0$ and $\eta \simeq 0.9$ this reduction is of the order of 80\%. These results demonstrate that the reduction of $Q_{\textrm{sc}}$ is robust against the variation of $B$. For $\eta \lesssim 0.9$ the increase of ${\bf B}$ not only improves the efficiency of the cloak, by further decreasing $Q_{sc}$ with respect to $Q_{sc}^{(0)}$, but also shifts the operation of the device to lower values of $\eta$. This result contrasts to the behavior of $Q_{sc}$ shown in Fig.~\ref{Figura3}a for different $\varepsilon_c$ and $\varepsilon_s$, where an increase in $B$ shifts the operation range of the cloak to higher values of $\eta$. This means that, for fixed $\eta$, the scattering signature of the system, and hence the cloaking operation functionality, can be drastically modified by varying either $B$ or the frequency ({\em i.e.}, by varying $\varepsilon_c$ and $\varepsilon_s$), demonstrating the versatility of the magneto-optical cloak.

In Fig.~\ref{Figura3}c, $Q_{sc}$ is calculated as a function of $\eta$ for $\varepsilon_c = 10\varepsilon_0$ and $\varepsilon_s = -1.5\varepsilon_0$, that are values of permittivity that preclude the possibility of invisibility for any $\eta$ and ${\bf B} = {\bf 0}$. Figure \ref{Figura3}c reveals that for ${\bf B} \neq {\bf 0}$ a very strong reduction in $Q_{sc}$ can occur for a wide range of values of $\eta$, showing that one can drastically reduce the scattered field by applying a magnetic field to a system that is not originally designed to perform as an EM cloak. Indeed, for $\gamma_s = 3\varepsilon_0, 5\varepsilon_0, 12\varepsilon_0$ the reduction in $Q_{sc}$ ranges from $80\%$ to $95\%$ for $0 < \eta \leq 0.5$. This result complements the one depicted in Fig.~\ref{Figura3}b, where the presence of ${\bf B}$ was also shown to be capable to switch off the functionality of the system as a cloak. The crossover between these distinct scattering patterns could be achieved by varying the incident wave frequency.

Figure~\ref{Figura3}d also corresponds to the case where invisibility cannot occur for ${\bf B} = {\bf 0}$ ($\varepsilon_c = 10\varepsilon_0$ and $\varepsilon_s = 1.5\varepsilon_0$). For ${\bf B} \neq {\bf 0}$ again there is a reduction of $Q_{sc}$ for $ \eta > 0.8$, which can be of the order of $50\%$ for $\gamma_s = 12\varepsilon_0$, further confirming that the application of ${\bf B}$ to a cylinder coated with a magneto-optical shell can
switch on the functionality of the device as an invisibility cloak.

\begin{figure}
\centering
\includegraphics[scale=0.42]{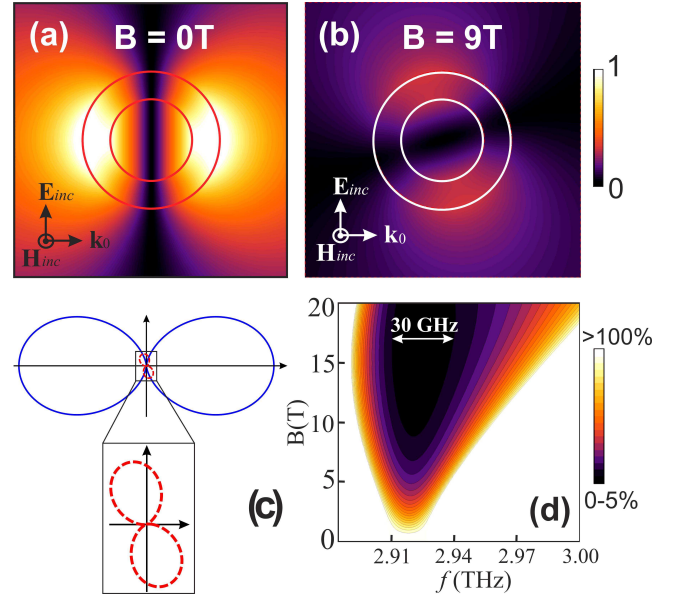}
\caption{Spatial distribution of the scattered field $H_{z}$ for $B=0$ (a) and (b) $B=9\,T$ for $\varepsilon_c = 10\varepsilon_0$. The incident frequency is $\omega/2\pi = 2.93$THz and $a/b= 0.6$ with $b = 0.01\lambda$. Differential scattering cross-section corresponding to the cases in (a) (blue solid curve) and (b) (dashed red curve) is shown in panel (c). (d) Scattering efficiency (normalized by $Q_{sc}^{(0)}$) as a function of the frequency and $B$.
}\label{Figura4}
\end{figure}

To discuss a possible experimental realization of the system proposed here, let us consider the existing cylindrical implementation of a cloaking device described in Ref.~\cite{rainwater2012}, but without the parallel-plate implants. The shell is assumed to be made of a magneto-optical dielectric material described by the Drude-Lorentz model~\cite{King2009} with realistic material parameters, including losses: oscillating strength frequency $\Omega/2\pi = 3$THz, resonance frequency $\omega_0/2\pi = 1.5$THz, and damping frequency $\Gamma/2\pi = 0.03$THz. Drude-Lorentz permittivities are extensivelly used to describe a wide range of media, including magneto-optical materials. A particularly recent example is monolayer graphene epitaxially grown on SiC~\cite{crassee2012}, where the Faraday rotation is very well described by a anisotropic Drude-Lorentz model, with similar parameters to the ones proposed here.
For concreteness, we set the device to operate around the frequency $f = 2.93$ THz; the inner and outer radii are $a = 0.6b$ and $b = 0.01\lambda$, respectively. The core is made of a dielectric with $\varepsilon_c = 10\varepsilon_0$, and negligible losses. The spatial distribution of the scattered field $H_z$ in the $xy$ plane is shown in Fig. \ref{Figura4}. For this set of parameters invisibility cannot occur for ${\bf B}= {\bf 0}$, and the corresponding spatial distribution of the scattered field is dipole-like (Fig. \ref{Figura4}a), as expected since for $b\ll \lambda$ the electric dipole term is dominating. Figure \ref{Figura4}b reveals that the presence of ${\bf B}$ strongly suppresses the scattered field for all angles, further indicating that ${\bf B}$ could play the role of an external agent to switch on and off the cloaking device. This result is corroborated by Fig. \ref{Figura4}c, where the differential scattering efficiency is calculated without and with the presence of ${\bf B}$. Figure \ref{Figura4}c confirms that the scattered radiation, which is initially dipole-like for ${\bf B}= {\bf 0}$, is strongly suppressed in all directions when ${\bf B}$ is applied, even in the presence of material losses. Figure \ref{Figura4}d exhibits a contour plot of the scattering efficiency $Q_{sc}$ (normalized by $Q_{sc}^{(0)}$) as a function of both $B$ and the incident wave frequency $f$. From Fig. \ref{Figura4}d one can see that the reduction of $Q_{\textrm{sc}}$ can be as large as $95$\% if compared to the case without ${\bf B}$.
We emphasize that these findings are robust against material losses, illustrated by the fact that we are allowing for quite typical dissipation parameters and the cloak works at the levels discussed above. Furthermore, we checked that even for much larger dissipative systems, characterized by $b=0.1\lambda$ and $b=0.2\lambda$, the cross section reduction (calculated including up to the electric octopole contribution) can be as impressive as $92\%$ and $83\% $, respectively. In addition, our results suggest that the efficiency of the proposed system is comparable to state-of-the-art existing cloaking apparatuses~\cite{edwards2009,rainwater2012} with the advantage of being highly tunable in the presence of magnetic fields. Besides, Fig. \ref{Figura4}d demonstrates that for  $B \simeq 15 T$ the reduction of $Q_{\textrm{sc}}$ of the order of $95$\% occurs for a relatively broad band of frequencies, of the order of 30 GHz. Finally, the effect of increasing $B$ around the design operation frequency is to broaden the frequency band for which the reduction of $Q_{\textrm{sc}}$ induced by the magnetic field is as large as $95$\%.

There are dielectric materials that exhibit strong magneto-optical activity that could be used in the design of a magneto-optical cloak. Single- and multilayer graphene are promising candidates, as they exhibit giant Faraday rotations for moderate magnetic fields~\cite{crassee2011}. Garnets are paramagnetic materials with huge magneto-optical response, with Verdet constants as high as $10^{4}$ deg./[T.m.] for visible and infrared frequencies~\cite{barnes1992}. There are composite materials made of granular magneto-optical inclusions that show large values of $\gamma_s$ for selected frequencies and fields in the 10-100 T range~\cite{Stroud1990, Reynet2002}. These materials offer an additional possibility of tuning EM scattering by varying the concentration of inclusions, and have been successfully employed in plasmonic cloaks~\cite{farhat2011}. It is worth mentioning that the magneto-optical response of typical materials, which will ultimately govern the tuning speed of the system, is usually very fast. Indeed, magneto-optical effects manifest themselves in a time scale related to the spin precession; for typical paramagnetic materials this time is of the order of nanoseconds~\cite{kirilyuk2010}. Hence the tuning mechanism induced by magneto-optical activity is expected to be almost instantaneous after the application of ${\bf B}$.

In conclusion, we investigate EM scattering by a dielectric cylinder coated with a magneto-optical shell to conclude that one could actively tune the operation of plasmonic cloaks with an external magnetic field ${\bf B}$. In the long wavelength limit we show that the application of ${\bf B}$ may drastically reduce the scattering cross-section for all observation angles. The presence of ${\bf B}$ can also largely modify the operation range of the proposed magneto-optical cloak in a dynamical way. Indeed, for a fixed $\eta$, we demonstrate that invisibility can be achieved by applying a magnetic field for situations where the condition for transparency cannot be satisfied without ${\bf B}$. Conversely, a magnetic field can suppress invisibility in a system originally designed to act as an invisibility cloak. Together, these results suggest that one can dynamically switch on and off the magneto-optical cloak by applying ${\bf B}$. In terms of frequency, the application of ${\bf B}$ in a system with fixed $\eta$ could largely shift the cloak operation range, both to higher and lower frequencies. We also show that these results are robust against material losses and discuss the feasibility of designing a magneto-optical cloak using existing materials and moderate magnetic fields. We hope that our results could guide the design of dynamically tunable, versatile plasmonic cloaks, and optical sensors.

%
%

We thank R. M. de Souza, V. Barthem, and D. Givord for useful discussions, and FAPERJ, CNPq and CAPES for financial support.


\end{document}